\documentclass[aps,prl,twocolumn,amsmath,amssymb,showpacs]{revtex4-1}

\usepackage[ansinew]{inputenc}
\usepackage{graphicx}
\usepackage{textcomp}
\usepackage{amsmath}
\usepackage{amssymb}

\begin{document}

\title{Erbium dopants in silicon nanophotonic waveguides}
\author{Lorenz Weiss}
\email{L.W. and A.G. contributed equally to this work.}
\author{Andreas Gritsch}
\email{L.W. and A.G. contributed equally to this work.}
\author{Benjamin Merkel}
\author{Andreas Reiserer}
\email{andreas.reiserer@mpq.mpg.de}

\affiliation{Quantum Networks Group, Max-Planck-Institut f\"ur Quantenoptik, Hans-Kopfermann-Strasse 1, D-85748 Garching, Germany}
\affiliation{Munich Center for Quantum Science and Technology (MCQST), Ludwig-Maximilians-Universit\"at
M\"unchen, Fakult\"at f\"ur Physik, Schellingstr. 4, D-80799 M\"unchen, Germany}

\begin{abstract}
The combination of established nanofabrication with attractive material properties makes silicon a promising material for quantum technologies, where implanted dopants serve as qubits with high density and excellent coherence even at elevated temperatures \cite{morton_embracing_2011,  vandersypen_interfacing_2017}. In order to connect and control these qubits, interfacing them with light \cite{yin_optical_2013,  morse_photonic_2017, chartrand_highly_2018, awschalom_quantum_2018} in nanophotonic waveguides offers unique promise \cite{wehner_quantum_2018, chang_colloquium_2018}. Here, we present resonant spectroscopy of implanted erbium dopants in such waveguides. We overcome the requirement of high doping and above-bandgap excitation that limited earlier studies \cite{przybylinska_optically_1996, kenyon_erbium_2005, vinh_photonic_2009, yin_optical_2013}. We thus observe erbium incorporation at well-defined lattice sites with a thousandfold reduced inhomogeneous broadening of about $1\,\text{GHz}$ and a spectral diffusion linewidth down to $45\,\text{MHz}$. Our study thus introduces a novel materials platform for the implementation of on-chip quantum memories, microwave-to-optical conversion, and distributed quantum information processing, with the unique feature of operation in the main wavelength band of fiber-optic communication.
\end{abstract}

\maketitle


Individual dopants and other atom-like defects in solids are promising candidates for the realization of quantum computers \cite{morton_embracing_2011, vandersypen_interfacing_2017} and quantum networks \cite{awschalom_quantum_2018, wehner_quantum_2018}. Among all optically active dopants studied to date, either in silicon \cite{morse_photonic_2017, chartrand_highly_2018} or in other crystals \cite{awschalom_quantum_2018, liu_spectroscopic_2005}, erbium stands out because its emission wavelength falls within the main wavelength band of optical telecommunication between $1530\,\text{nm}$ and $1565\,\text{nm}$. The transparency of silicon in this wavelength regime ensures compatibility with the mature platform of silicon nano-photonics \cite{vivien_handbook_2013}. In addition, the minimal loss of optical fibers at this wavelength is a key requirement for quantum networks that span global distances.

In this context, the coherence time of the quantum state of individual dopants is paramount. For erbium in silicate crystals, the coherence of the ground state transition can exceed one second in high magnetic fields \cite{rancic_coherence_2018}, and the $4\,\text{ms}$ long coherence of the optical transition \cite{liu_spectroscopic_2005} is the best measured in any solid. The reason for this exceptional coherence is the confinement of the inner 4f electrons near the erbium nucleus and the shielding of the crystal field by the outer 5s and 5p electrons. This protects the electronic state from decoherence via phonons \cite{liu_spectroscopic_2005}, and also hinders phonon-sideband emission that reduces the radiative efficiency of other defects \cite{awschalom_quantum_2018, morse_photonic_2017, chartrand_highly_2018}.

The excellent coherence of erbium dopants has enabled first experiments towards quantum information processing, including efficient memory for both optical \cite{dajczgewand_large_2014, saglamyurek_multiplexed_2016} and microwave photons \cite{probst_microwave_2015}, and the electrical detection of individual dopants in silicon \cite{yin_optical_2013}. In order to take full advantage of this platform, the integration into nanophotonic structures is highly desirable. This not only allows for efficient multiplexing via robust and cost-effective fabrication of devices, but can also enhance the interaction strength between the dopants and light fields at the single photon level \cite{lodahl_interfacing_2015}, thus overcoming the small oscillator strength of the erbium transition \cite{liu_spectroscopic_2005}.

Pioneering experiments in this direction have used different host materials, including potassium titanyl phosphate, lithium niobate and yttrium orthosilicate \cite{thiel_rare-earth-doped_2012, miyazono_coupling_2016,  dibos_atomic_2018, raha_optical_2020, askarani_persistent_2020}. All of these materials are incompatible with standard Complementary Metal Oxide Semiconductor (CMOS) processing and have abundant isotopes with nuclear spins, such that superhyperfine couplings limit the achievable coherence time both in ground \cite{raha_optical_2020} and excited states \cite{car_selective_2018}. In contrast, silicon is a predominantly spin-free material. Further isotopic purification completely eliminates the coupling of dopants to nuclear spins \cite{saeedi_room-temperature_2013} and facilitates ultra-narrow optical linewidths \cite{chartrand_highly_2018}. This makes the direct integration of erbium dopants into silicon nanophotonic structures, as studied in this work, highly promising. 

\begin{figure*}
\includegraphics[width=2.0\columnwidth]{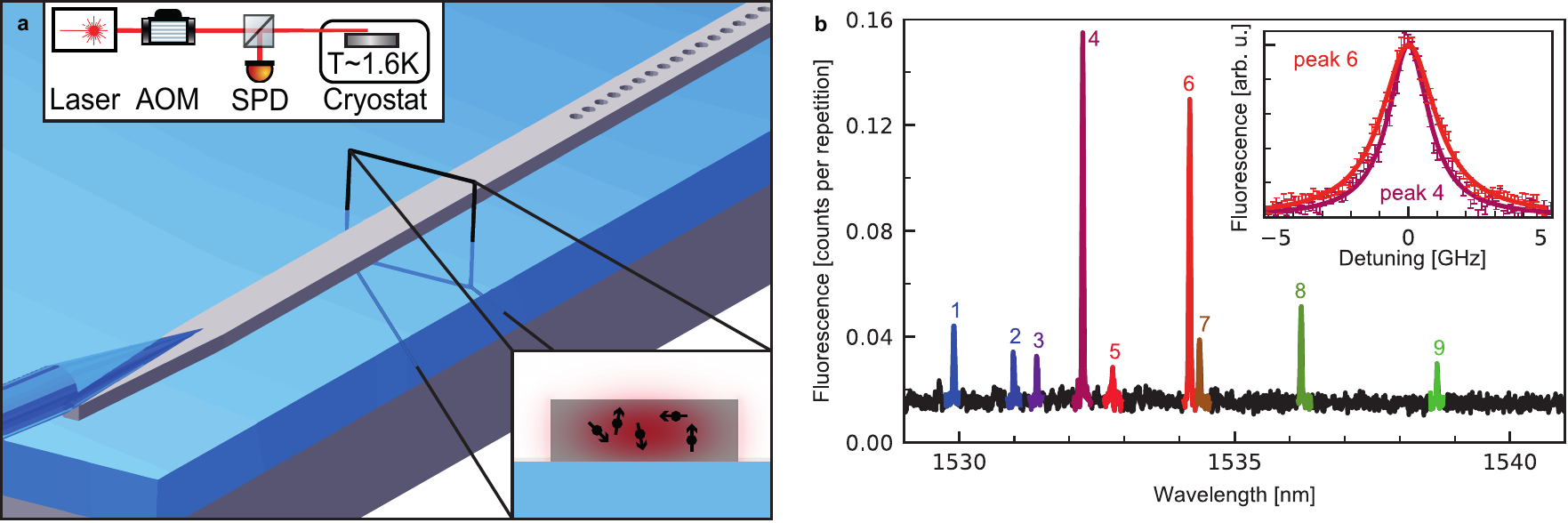}
\caption{        \label{fig:setup_spectrum}
\textbf{ a. Experimental setup} (not to scale). We use a silicon nanowire waveguide (grey) with a cross section of $220\,\times\, 700\,\text{nm}$, terminated by a broadband photonic crystal mirror (dark grey holes). The devices are fabricated on a commercial silicon-on-insulator chip (blue). Efficient broadband coupling is achieved with a tapered fiber touching the waveguide (blue). The bottom right inset shows a cross section. Erbium (black arrows) has been implanted with an energy of 198 keV, which results in an approximately Gaussian distribution centered within the fundamental mode (red) of the waveguide. The top inset shows a sketch of the experimental setup. The sample is mounted on a nanopositioning stage and kept at a temperature of 1.6 K in a closed-cycle cryostat. A tunable laser is pulsed using two acousto-optical modulators (AOM). Using a beam splitter, the emitted fluorescence is detected using a single-photon detector (SPD).
\textbf{ b. Fluorescence measurement}. We observe nine narrow fluorescence peaks in the telecom C-band. The laser frequency is swept during the scan to avoid effects of persistent spectral hole burning. In the absence of non-radiative decay, the peak height is thus proportional to the concentration of erbium at a given site. All observed peaks are well fit by Lorentzian curves. The linewidths range from 1 to $4\,\text{GHz}$. The inset shows the two strongest peaks (labeled 4 and 6; the other peaks are shown in Figure \ref{fig:AllData}). 
}
\end{figure*}

Unfortunately, the low solubility of erbium in silicon prevents the incorporation of a significant density of erbium dopants during crystal growth from the melt \cite{kenyon_erbium_2005}. Therefore, spectroscopy requires samples prepared by nonequilibrium techniques, such as epitaxial growth or ion implantation \cite{przybylinska_optically_1996}. The resulting two-dimensional samples give only a small absolute number of Er spins in the focal spot of an objective. In addition, the high refractive index of silicon ($n \simeq 3.5$) leads to a small critical angle for total internal reflection, such that only about $0.1 \%$ of the erbium fluorescence can be collected from free space when using a planar sample with randomly oriented dipoles. Combined with the considerable inhomogeneous broadening of the emission and the long optical lifetime, exceeding 1 ms in all materials studied so far \cite{liu_spectroscopic_2005}, this has precluded resonant spectroscopy of few-ion ensembles.

Previous experiments, mostly in the context of lasers, instead used high doping concentration and above-bandgap excitation \cite{przybylinska_optically_1996, kenyon_erbium_2005, vinh_photonic_2009}. While the latter has the advantage of exciting many ions at the same time to provide a strong signal, it comes at the price of limiting the sensitivity to the small fraction of implanted dopants that is coupled to the conduction band \cite{kenyon_erbium_2005}. The same holds for recent experiments that used electro-optical detection \cite{yin_optical_2013, zhang_single_2019}.


In this work, we overcome this limitation by integrating erbium dopants into nanophotonic wire waveguides  \cite{vivien_handbook_2013}, as shown in figure \ref{fig:setup_spectrum}a.  With broadband side-coupling, we detect around 2\% of the photons emitted by the erbium dopants. Albeit this number can be improved by an order of magnitude in an optimized setup, the achieved detection efficiency allows us to perform pulsed resonant fluorescence spectroscopy in weakly doped samples using a $0.4\,\text{mm}$ long silicon waveguide. 

In a first experiment, we scan the frequency of the excitation laser and measure the fluorescence after the laser is switched off, using excitation pulses of $1\,\text{ms}$ duration. We sweep the laser frequency by a few hundred MHz during each excitation pulse to avoid  bleaching by persistent spectral holeburning to long-lived spin states, which has also been observed in other bulk crystals \cite{liu_spectroscopic_2005, rancic_coherence_2018, car_selective_2018, raha_optical_2020}, waveguides \cite{thiel_rare-earth-doped_2012, askarani_persistent_2020} and even optical fibers \cite{saglamyurek_multiplexed_2016}.

In contrast to all previous measurements \cite{przybylinska_optically_1996, kenyon_erbium_2005,vinh_photonic_2009, yin_optical_2013, zhang_single_2019}, our resonant fluorescence technique is sensitive to all dopants that decay radiatively. We therefore observe a completely different spectrum,  shown in figure \ref{fig:setup_spectrum}b, which exhibits several sharp Lorentzian peaks, as expected for crystals with low defect concentration \cite{thiel_rare-earth-doped_2012}. Their inhomogeneous linewidth is three orders of magnitude smaller than reported previously for erbium-implanted silicon \cite{przybylinska_optically_1996, kenyon_erbium_2005}, ranging from 1 to $4\,\text{GHz}$ FWHM.

The observation of such narrow resonances indicates that our fabrication procedure (detailed in the methods section) achieves dopant integration at well-defined lattice sites \cite{liu_spectroscopic_2005}, which is a key step towards quantum controlled applications of erbium-doped silicon. It is not only beneficial for achieving a sufficient optical depth for on-chip quantum memories, but also indicates that our implantation and annealing procedure does not substantially damage the crystal structure, which would lead to a large inhomogeneous broadening and a short coherence time of embedded rare-earth dopants \cite{liu_spectroscopic_2005}.

Similar to other host crystals, the remaining inhomogeneous broadening is attributed to random strain fields in the waveguide that are caused by the different thermal expansion coefficients of Si and $\mathrm{SiO}_2$, by the considerable amount of oxygen and other impurities in commercial silicon-on-insulator wafers, and, finally, by the mixed isotopic composition of the crystal. We expect that similar to other dopants and emitters in silicon, these sources can be largely eliminated in optimized samples \cite{saeedi_room-temperature_2013, chartrand_highly_2018}, opening exciting prospects for ultra-narrow optical lines in nanofabricated waveguide and resonator structures.

\begin{figure}
\includegraphics[width=1.\columnwidth]{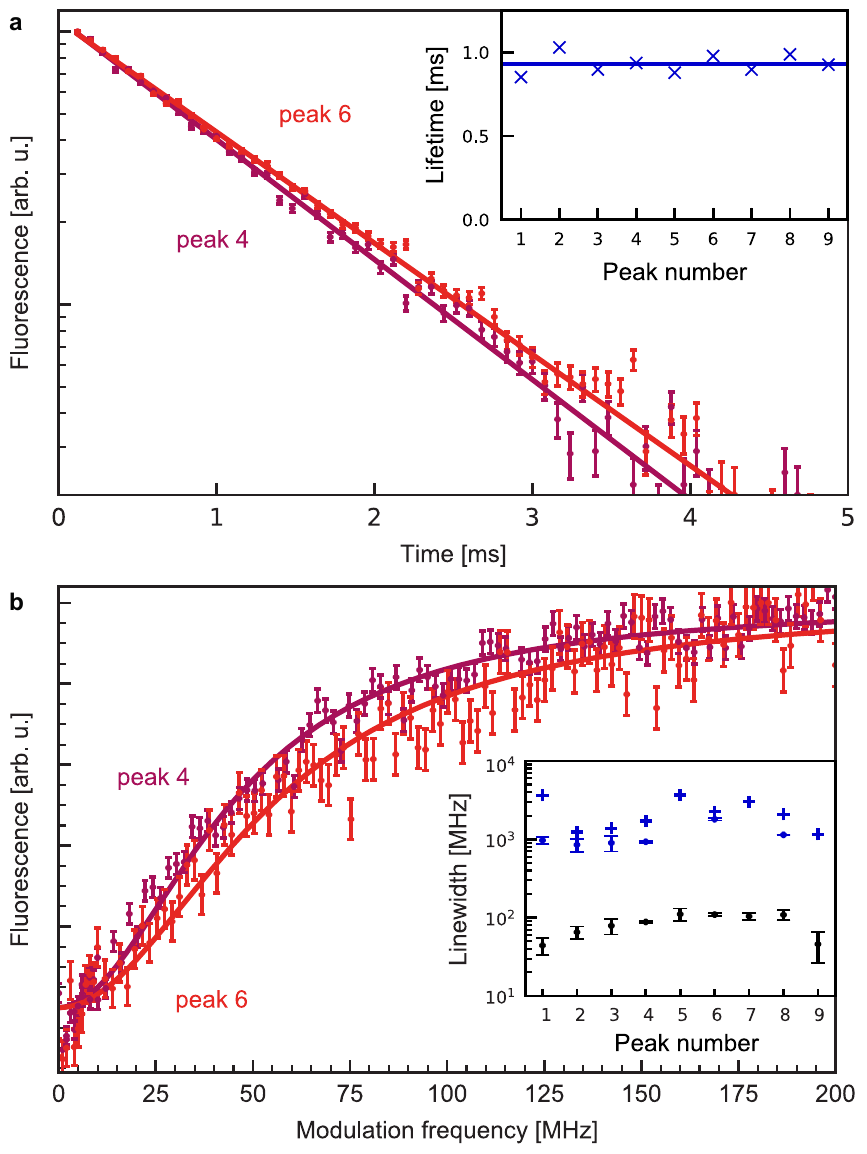}
\caption{ \label{fig:Lifetime_Linewidth}
\textbf{a. Optical lifetime}. After the laser is switched off, the fluorescence decays exponentially with time, shown for the two strongest peaks (4 and 6). Inset: The fitted lifetime (blue data points) of the individual peaks is close to the mean value (blue line).
\textbf{b. Spectral diffusion linewidth}. When the laser is modulated with a frequency that is larger than the spectral diffusion linewidth of the dopants, saturation is reduced and the fluorescence increases. The data is well fit by inverted Lorentzian curves, shown for the two strongest fluorescence lines (4 and 6). The other peaks are displayed in Figure \ref{fig:AllData}. Inset: The FWHM of the fit (black) depends on the site. Comparing to the inhomogeneous broadening FWHM without (blue crosses) or with (blue dots) applied magnetic field yields a ten- to hundredfold narrower spectral diffusion linewidth.
}
\end{figure}


After characterizing the inhomogeneous broadening of the emitters, we investigate their lifetime. To this end, we recorded the temporal decay of the fluorescence after the excitation laser is switched off. As shown in figure \ref{fig:Lifetime_Linewidth}a, we observe an exponential decay with a lifetime around 1 ms for every peak (as shown in the inset). 

Compared to other established host materials for erbium \cite{liu_spectroscopic_2005}, the excited state lifetime is reduced by about an order of magnitude, which holds promise for an enhanced light-matter interaction strength. Still, the lifetime is much larger than that of other solid-state quantum emitters that typically lies in the range of ten nanoseconds \cite{awschalom_quantum_2018}. The reason is that transitions between the 4f shells of rare-earth ions in sites of high symmetry are forbidden by parity. However, this does not hold for transitions via magnetic dipole coupling to the radiation field. The high refractive index of silicon lets us expect a purely magnetic lifetime of only 1.5 ms \cite{dodson_magnetic_2012}, close to the measured value. We therefore conclude that the character of the decay is predominantly radiative and via magnetic dipole transitions, in agreement with the observation that the lifetime is almost independent of the sites.

The slight deviation between the measurement and the theoretically expected value could be caused by forced electric dipole transitions, by inaccuracies in the calculation, by weak nonradiative decay, and by the small modification of the electromagnetic density of states in nanowire waveguides \cite{stepanov_quantum_2015}. Tailored structures, such as photonic crystal waveguides and cavities, can be used to further shorten the lifetime if desired \cite{lodahl_interfacing_2015}. But even in this case, using erbium-doped silicon in advanced quantum networking protocols \cite{reiserer_cavity-based_2015, wehner_quantum_2018}  will require small homogeneous and spectral diffusion linewidths, which we characterize in the following via transient spectral hole burning \cite{liu_spectroscopic_2005}.

Our measurement technique is based on the observation that the fluorescence signal $S$ increases nonlinearly with the laser intensity $I$ applied at a single frequency, $S \propto \sqrt{I}$, because of saturation effects. We therefore apply three laser fields of about the same intensity $I$, generated at equidistant frequency separation within the inhomogeneous linewidth by modulating the input laser field with an electro-optical modulator.

If the separation of the three laser lines is larger than both the homogeneous and the spectral diffusion linewidth of the ions, irradiating them will increase the fluorescence threefold compared to that of a single field. If, however, the detuning is small, or zero, the fluorescence will only increase by a factor of about $\sqrt{3}$. Thus, by scanning the modulation frequency, we can measure an upper bound to the homogeneous and spectral diffusion linewidth of the excited erbium dopants on the timescale of their radiative lifetime when using excitation pulses of 1 ms duration. Again, to preclude detrimental effects of persistent spectral hole burning, the laser frequency is changed by a few MHz between different repetitions of the experiment. The measurement results for the two strongest lines are shown in figure \ref{fig:Lifetime_Linewidth}b.

The data are fit with inverted Lorentzian curves, resulting in a spectral diffusion linewidth between $45$ and $110\,\text{MHz}$ for the individual sites, as shown in the inset (black data). The observed value for erbium dopants in silicon is thus of the same order as in other hosts \cite{liu_spectroscopic_2005}. The observed broadening can have two major causes. We can exclude the first, dipolar interactions with other magnetic moments in the crystal, as both our dopant concentration and the interaction with the nuclear spin bath is too small.

The second cause of spectral diffusion can be crystalline defects and the proximity of interfaces, a common issue for all solid-state quantum emitters \cite{awschalom_quantum_2018, lodahl_interfacing_2015} including erbium in other hosts \cite{thiel_rare-earth-doped_2012, dibos_atomic_2018, askarani_persistent_2020}. At the used intensity, we expect to generate on the order of $10^4$ free carriers in the waveguide during each excitation pulse via two-photon absorption \cite{vivien_handbook_2013}. This can change the state of charge traps caused by crystalline defects and dangling bonds at the surface, leading to fluctuating electric fields at the position of the dopant and thus a broadening of the line via the Stark effect. In our samples, the maximum distance of the dopants to the closest interface is around $100\,\text{nm}$. With typical Stark coefficients of $100\,\mathrm{Hz}\cdot\mathrm{m}/\mathrm{V}$ \cite{zhang_single_2019}, the expected frequency shift for single fluctuating charge trap at this distance is tens of MHz, matching our observations. 

In future devices, spectral diffusion may therefore be reduced by using waveguides of larger dimension, lower excitation power, and surface termination (e.g. by hydrogenation). In addition, stabilizing the state of the charge traps by applying electric fields, as demonstrated recently in silicon carbide \cite{anderson_electrical_2019}, seems highly promising.


\begin{figure}
\includegraphics[width=1.0\columnwidth]{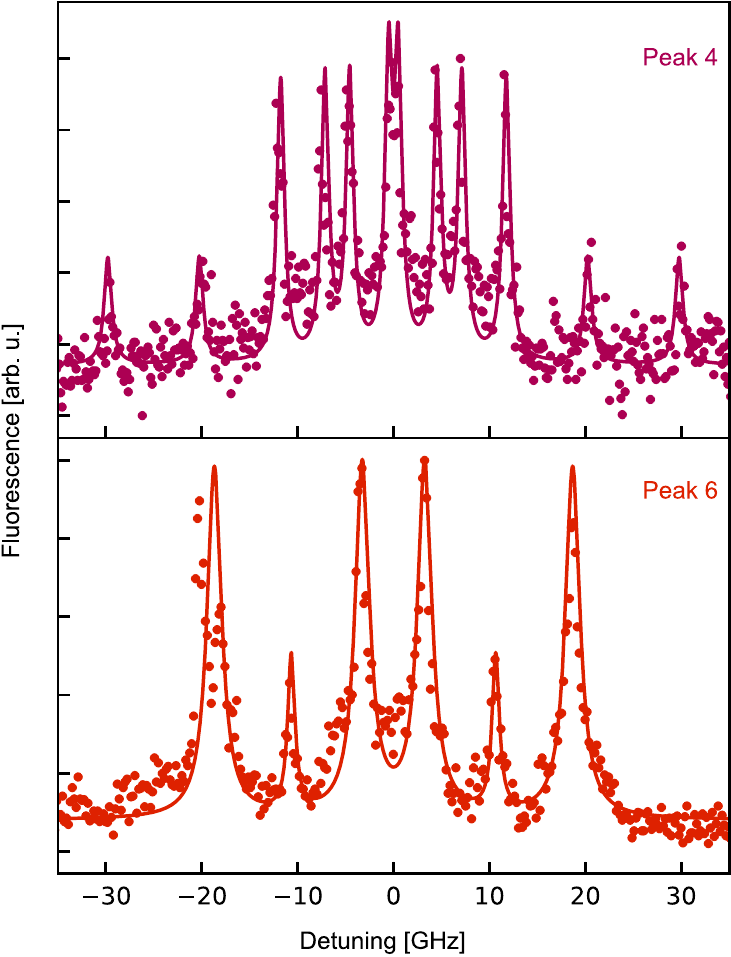}
\caption{\label{fig:BField}
\textbf{Magnetic field dependence}. The degeneracy of the erbium ground and excited state doublets is lifted by applying a magnetic field of $0.2\,\text{T}$ along the [100] axis of the crystal. The observed splitting differs between the lines, shown here for the strongest peaks and in Figure \ref{fig:AllData} for the others. The different height of the individual lines indicates a different transition strength for spin-preserving and spin-flip transitions. Peaks that show a larger number of split lines correspond to erbium sites with a lower symmetry in the silicon lattice. The data is fit to several Lorentzian lines with the same FWHM, which gives a smaller inhomogeneous broadening than observed without magnetic field.
}
\end{figure}

\begin{figure*} [h]
\includegraphics[width=1.9\columnwidth]{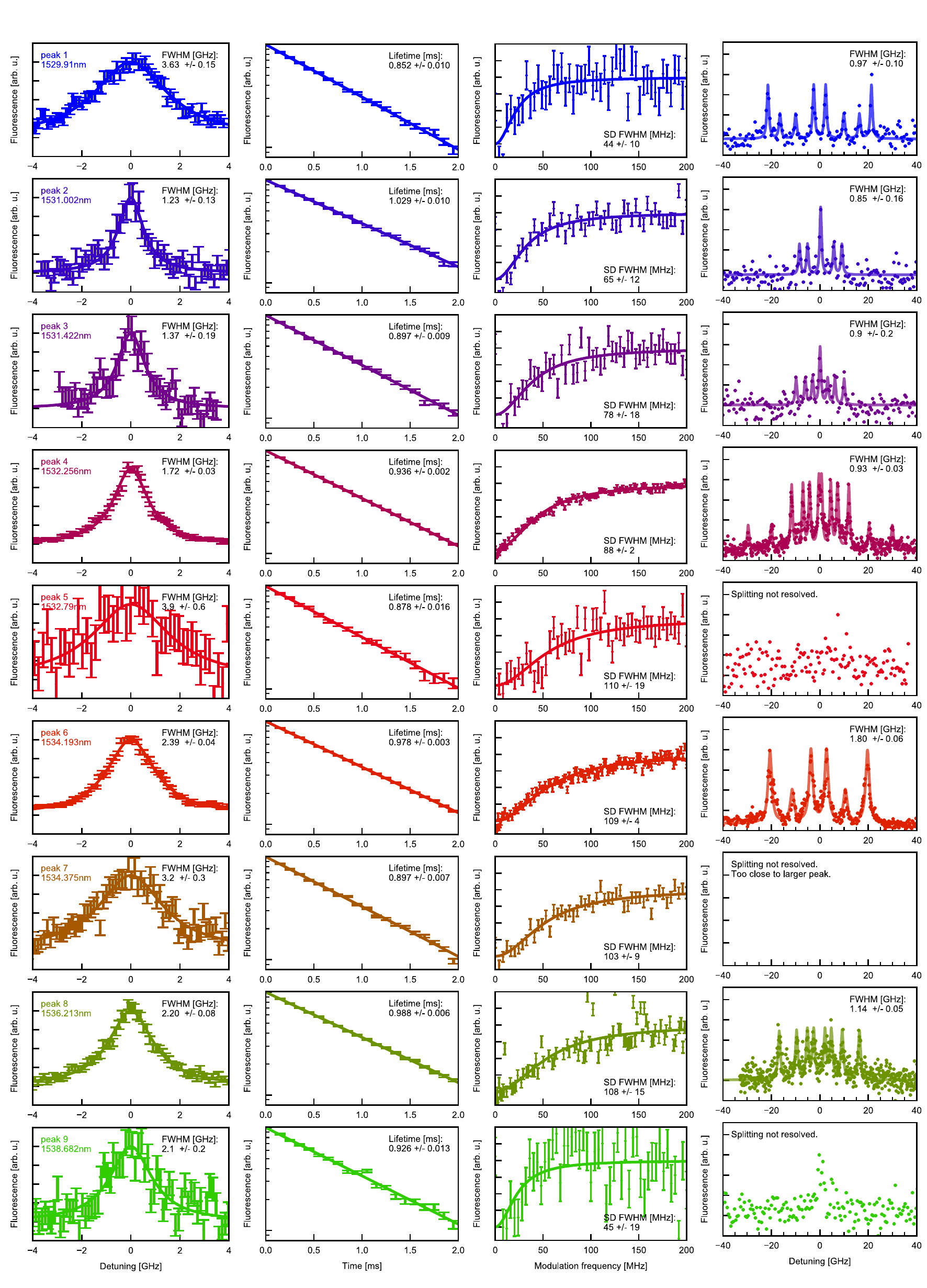}
\caption{     \label{fig:AllData}
\textbf{Direct comparison of all peaks (top to bottom).}
From left to right: Inhomogeneous linewidth, lifetime, spectral diffusion linewidth, and splitting in a magnetic field of $0.2\,\text{T}$.
}
\end{figure*}

After characterizing the lifetime and linewidth of the different erbium sites, we finally investigate their crystallographic properties. As erbium is a Kramer's ion \cite{liu_spectroscopic_2005}, every energy level will show a two-fold degeneracy that is lifted when applying a magnetic field. For a single dopant occupying a site with cubic symmetry, the optical transitions will therefore split into four separate lines. For sites of lower symmetries, even more lines are expected.
Figure \ref{fig:BField} shows the spectra observed for the two strongest lines, which clearly differ in symmetry. The splitting of the outermost lines is on the order of 20 GHz. This indicates a large difference between the effective g-factors of the ground- and excited state, on the order of 10, which is typical for erbium in various hosts \cite{liu_spectroscopic_2005}. The spectra of the other fluorescence lines are shown in Figure \ref{fig:AllData}. Determining the exact crystallographic properties of each of the sites requires a rotation of the magnetic field, which is left for future work. 


In summary, by integrating ensembles of erbium dopants with narrow linewidth into nanophotonic waveguides we have introduced a novel materials platform for quantum technology. Well-established nanofabrication techniques will allow for the realization of robust, low-cost, and multiplexed quantum devices. The observed narrow inhomogeneous linewidths may be used for on-chip laser stabilization. The strong coupling to magnetic fields, enabled by the high effective g factors, might allow for efficient quantum microwave-to-telecom conversion in nanofabricated waveguides \cite{obrien_interfacing_2014}.

In addition, our system can be used for spatially multiplexed, on-chip quantum memories. We expect that the second-long hyperfine coherence time demonstrated recently with erbium-doped silicate crystals \cite{rancic_coherence_2018} can also be achieved in silicon. Still, implementing a quantum memory will require a higher optical depth than used in this work. We expect that a hundredfold increase of the latter can be achieved by optimizing the implantation and annealing procedure to give a homogeneous dopant distribution with a high concentration of dopants in a single fluorescence line. Further, a hundredfold increase of the waveguide length with less than 10 \% loss is possible using nanophotonic rib waveguides \cite{vivien_handbook_2013}. As an alternative, the light-matter interaction can be strongly enhanced via collective effects in nanophotonic waveguides \cite{chang_colloquium_2018}, and via optical resonators \cite{lodahl_interfacing_2015}. Using ultra-high Q photonic crystal cavities, as demonstrated recently \cite{asano_photonic_2018}, a Purcell-enhancement by six orders of magnitude is expected. This would not only bring the system into the strong-coupling regime of cavity QED \cite{reiserer_cavity-based_2015}, but also shorten the radiative decay to the nanosecond range, such that the lifetime-limited linewidth is of the same order as the spectral diffusion linewidth we observed. Thus, we expect that erbium-doped silicon can be used as an optical interface of single spin qubits \cite{morton_embracing_2011,morse_photonic_2017}, operating in the telecom C-band. This offers unique promise for cavity-based quantum networks \cite{reiserer_cavity-based_2015, wehner_quantum_2018} and distributed quantum information processors based on a scalable platform.

\section{Acknowledgements}
We acknowledge measurement assistance by Tangui Aladjidi and Zarije Ademi as well as helpful discussions with Hannes Bernien, Ralf Riedinger and Thierry Chaneli\`ere. This project received funding from the European Research Council (ERC) under the European Union's Horizon 2020 research and innovation programme (grant agreement No 757772), from the Deutsche Forschungsgemeinschaft (DFG, German Research Foundation) under Germany's Excellence Strategy - EXC-2111 - 390814868, and from the Daimler-and-Benz-Foundation.

\bibliographystyle{naturemagsr.bst}
\bibliography{bibliography.bib}

\section{Supporting information}

\subsection{Sample preparation}

The samples are fabricated from a commercial weakly P-doped (10 ${\Omega}{\cdot}$cm), Czochralski-grown silicon-on-insulator wafer with 220 nm device and 3 micron buried-oxide layer thickness (Silicon Valley Microelectronics). After dicing, the wafers are implanted with a  mixture of erbium isotopes at ambient temperature with an energy of 198 keV (Fraunhofer IISB), followed by an annealing step at $10^3$ K. Numerical simulations (SRIM.org) lead to an expected range of 76 nm and straggle of 16 nm. Sparking in the ion source led to a reduced dose compared to the target value of $10^{14}\, \mathrm{cm}^{-2}$. We therefore used secondary ion mass spectrometry to compare the dopant concentration to a reference sample of pure epitaxially grown silicon that has been implanted by a different supplier to $1.5\cdot10^{12}\, \mathrm{cm}^{-2}$. In this way, we estimate that the dose of the sample used for all measurements is below $10^{11}\, \mathrm{cm}^{-2}$.

After dicing, we fabricate photonic nanostructures by electron beam lithography using ZEP resist on a nB5 machine (Nanobeam Ltd.). The pattern is transferred via reactive ion etching (Oxford Plasmalab 80) using a mixture of CF4 and SF6 etch gases. The end of the silicon waveguide is terminated with a broadband photonic crystal reflector in order to collect light emitted into both waveguide directions.

\subsection{Experimental setup}
The samples are mounted on a nanopositioning platform (Attocube) into a closed cycle cryostat (Attocube Attodry 2100) and cooled in helium buffer gas to a temperature below 1.7 K. The erbium dopants are excited with laser pulses generated from a continuous-wave source (Toptica CTL 1550)  via two fiber coupled acousto-optical modulators (Gooch and Housego). The frequency of the laser is determined to a precision of 100 MHz using a calibrated spectrum analyzer (Bristol instruments). For sweeping the frequency, the angle of the grating in the laser cavity is scanned. Without this scan, the fluorescence is reduced, for some peaks to an undetectable level, which we attribute to persistent spectral hole burning. The fluorescence signal is detected with a superconducting nanowire single photon detection system with a detection efficiency around $80\,\%$ and less than $10\,\text{Hz}$ dark count rate (Photon Spot).

To maximize the signal and thus be sensitive to small erbium dopant ensembles, we need a highly efficient and broadband off-chip coupling, which is best achieved in a side-coupling geometry \cite{vivien_handbook_2013}. To this end, we terminate the waveguide in an inverted taper and couple to it via a tapered single-mode fiber, fabricated by etching in HF. While this allows for a coupling efficiency approaching unity for free-standing structures, we observe coupling of 5 \% for structures that reside on the $\mathrm{SiO}_2$ buried oxide layer.

In this way, we expect to detect around 2 \% of the photons emitted by the erbium dopants. In this estimate we have used the independently measured optical losses (20 \%), detector efficiency (80 \%), and the calculated efficiency of collecting an emitted photon into the waveguide (about 40 \% for a isotroptic dipole and more than 90 \% for a transversal optical dipole located at the center of the waveguide \cite{stepanov_quantum_2015}).

\end{document}